\documentstyle[11pt]{article}
\newcommand{\be}{\begin{eqnarray}}

\newcommand{\ee}{\end{eqnarray}}
\newcommand{\munu}{\mu\nu}
\newcommand{\psibar}{\bar \psi}
\newcommand{\dirsh}{\!\!\!/}

\title{
        \begin{flushright}
        {\normalsize NBI--98--19\\
        August 1998 \\}
        \end{flushright}
\bf  Introduction to light cone field theory and high energy scattering}
\author{
        Raju Venugopalan \\
        {\small\it Niels Bohr Institute,
        Blegdamsvej 17,
        Copenhagen, Denmark, DK--2100 } \\
         }
\date{}

\begin{document}

\maketitle

\begin{center}
{\bf Abstract}\\
\end{center}
In this set of four lectures, we provide an 
elementary introduction to light cone
field theory and some of its applications in high energy scattering.

\vfill \eject

\section{Introduction}

In these lectures, we will attempt to provide a ``hands on'' introduction to 
some of the ideas and methods in light cone field theory and its 
application to high energy scattering. Light cone quantization as an approach
to study the Hamilton dymanics of fields was first investigated by Dirac, who 
pointed out several of its elegant features in a 
landmark paper~\cite{Dirac}. It was first applied to high energy physics 
in the 60's in the context of current algebra~\cite{Fubini}. Light cone field
theory currently finds applications in most areas of high energy physics, 
from perturbative QCD to string theories.

The elegance and simplicity of the light cone approach results from
the analogy of relativistic field theories quantized on the light cone
to non--relativistic quantum mechanics. In fact, this correspondence
runs deep and it was shown by Susskind that there is an exact
isomorphism between the Galilean subgroup of the Poincar\'{e} group
and the symmetry group of two dimensional quantum
mechanics~\cite{Susskind}.  Furthermore, as was first shown by
Weinberg~\cite{Weinberg}, the vacuum structure of field theories
simplifies greatly in the infinite momentum limit. The combination of
the non--relativistic kinematics of light cone field theories 
as well as their simple vacuum
structure, has given rise to the belief that potential methods of
quantum mechanics can be applied to field theories quantized on the
light cone.  This observation is at the heart of recent attempts to
understand bound state problems in QCD in the light cone
formalism~\cite{Wilson}. Indeed, beginning with the t'Hooft
model~\cite{tHooft} for mesons in 1+1--dimensional large $N_c$ QCD,
which made use of the light cone formalism, there have been many
attempts to study confinement and chiral symmetry breaking in this
approach (see Ref.~\cite{Matthias} and references therein).

Light cone field theory also provides much of the intellectual support for
the intuitive quark--parton picture of high energy scattering. Frequently, 
the phrases `the theory of strong interactions, QCD' and the 
`quark--parton picture of strong interactions' are used interchangeably.
However, it is only in light cone quantization (and light cone gauge) that
the quark--parton structure of QCD is manifest and multi--parton Fock states
can be constructed as eigenstates of the QCD Hamiltonian~\cite{Brodsky}. 
One can therefore construct Lorentz invariant light cone wavefunctions-- 
a fact which has been particularly useful in the study of exclusive processes
in QCD~\cite{BrodskyLepage}.
Further, in deeply inelastic scattering, the experimentally measured 
structure functions are simply related (in leading twist) to the light cone
quark distribution functions. The partonic picture of light cone quantum 
field theory was demonstrated very clearly in the papers of Kogut and 
Soper~\cite{KogutSoper} and of Bjorken, Kogut and Soper~\cite{BjKogutSoper}.

The goal of these lectures is to illustrate both of the above points, the
attractive features of light cone field theory and its applications to high
energy scattering, in the simplest possible fashion by working out concrete
examples. In the first lecture, we begin by introducing the light cone
notation and the two component formalism.  We then define the light cone Fock
states and their equal light cone time commutation relations. We conclude by
discussing the structure of the Poincar\'{e} group and demonstrate the above
mentioned isomorphism to two dimensional quantum mechanics.  In the second
lecture, we explicitly derive the light cone QCD Hamiltonian in the two
component formalism making use of the light cone constraint equations. It is
shown that the Hamiltonian can be expressed as the sum of non--interacting
and ``potential'' terms. For simplicity, in lecture three, we specialize to
the case of QED and use the form of the Hamiltonian derived in lecture 2 to
illustrate the parton picture of high energy scattering.  In particular, we
study high energy scattering off an external potential in the eikonal
approximation in QED. In the fourth and final lecture we show how Bjorken
scaling can be derived in QCD using the light cone commutation relations and
briefly discuss the relation of light cone distribution functions to
structure functions.

There are several reviews that the reader may study to learn more about the
subject. An introductory review which also includes a guide to the
literature for beginners is that by Harindranath~\cite{Avaroth}. Another 
introductory review which stresses recent advances is that by 
Burkardt~\cite{Matthias}. The most recent and comprehensive review of the
subject is by Brodsky, Pauli and Pinsky~\cite{BPP}. A part of our lectures 
relies 
heavily on the classic papers of Kogut and Soper~\cite{KogutSoper} and
Bjorken, Kogut and Soper~\cite{BjKogutSoper}. The reader should keep in 
mind that a wide variety of conventions are in use in the literature. Some
of these are discussed in the review of Brodsky, Pauli and Pinsky.

The lectures below were delivered at the Cape Town lecture school and
for spacetime reasons are the ``short'' form of lectures delivered
previously at the University of Jyv\"{a}skyl\"{a} international summer
school. The topics that were omitted in the short version include
light cone perturbation theory, the renormalization group and the
operator product expansion, and small x physics. The longer version of
these lectures will be published separately at a later date~\cite{Rajlater}.

\newpage

\section*{ Lecture 1: Light cone quantization and the light cone algebra.}
\vskip 0.15in

We begin by defining our convention and notations. Our metric here is the
$+2$ metric $\hat{g}^{\mu\nu}= (-,+,+,+)$. Note: for my convenience (and
unfortunately, your inconvenience) I may change notations in the latter
lectures. But you will have fair warning! The gamma matrices in usual
space--time co--ordinates are denoted by carets. In the chiral
representation,
\begin{displaymath}
{\hat{\gamma}}^0=\left( \begin{array}{ccc}
0 & I\\
I & 0\\
\end{array} \right)\,\, ; \,\,
{\hat{\gamma}}^i=\left( \begin{array}{ccc}
0 & \sigma^i\\
-\sigma^i & 0\\
\end{array} \right)\,\, ; \,\,
{\hat{\gamma}}^5=\left( \begin{array}{ccc}
I & 0\\
0 & -I\\
\end{array} \right)\,\, ,
\end{displaymath}
and $\{\hat{\gamma}^\mu,\hat{\gamma}^\nu\} = -2\hat{g}^{\mu\nu}$.
Above, $\sigma^i, i=
1,2,3$ are the usual $2\times2$ Pauli matrices and $I$ is the $2\times
2$
identity matrix. In light cone co--ordinates, $\gamma^{\pm}
= ({\hat{\gamma}}^0\pm {\hat{\gamma}}^3)/\sqrt{2}$ and
$\{\gamma^\mu,\gamma^\nu\}
=-2g^{\mu\nu}$, where $g^{++}=g^{--}=0$,  $g^{+-}=g^{-+}=-1$. Also, 
$g_{t_1,t_2}=1$
with $t_1,t_2=1,2$ denoting the two transverse co--ordinates.
We define $x^\mu \equiv(x^0,x^1,x^2,x^3)=(t,\vec{x})$ and
\be
x^\pm = {(t+z)\over{\sqrt{2}}}\,\,;\,\, \partial_\pm \equiv {\partial \over 
{\partial x^\pm}} = {1\over \sqrt{2}}(\partial_t \pm \partial_z)\,\,;\,\,
A^\pm = {(A^0 \pm A^z)\over \sqrt{2}} \, .
\ee
Note for
instance that in this convention $A_+ = -A^-$ and $A_t=+A^t$. Also, 
$q^2 = -2q^- q^+ + q_t^2$.  Hence, a ``space--like'' 
$q^2$ implying large space--like components would correspond to $q^2 >0$.

We now define the projection operators
\be
\alpha^\pm = {\hat{\gamma}^0\gamma^\pm \over \sqrt{2}} \equiv
{\gamma^\mp \gamma^\pm\over 2} \, ,
\ee
which project out the two component spinors $\psi_\pm = \alpha^\pm \psi$,
~\cite{KogutSoper}
\begin{equation}
\psi_+ =\left( \begin{array}{c}
0 \\
\psi_2\\
\psi_3\\
0\\ \end{array} \right)\,\, ; \,\,
\psi_- =\left( \begin{array}{c}
\psi_1 \\
0\\
0\\
\psi_4\\ \end{array} \right)\, ,
\end{equation}
where $\psi_1,\cdots \psi_4$ are the four components of $\psi$. 
It follows from the above that $\psi_+ + \psi_- = \psi$.

Some relevant properties of the projection operators $\alpha^\pm$ are
\be
(\alpha^\pm)^2 = \alpha^\pm \,\, ; \,\, \alpha^\pm\alpha^\mp =0 \,\, ;
\,\, \alpha^+ +\alpha^- =1 \,\, ; \,\, (\alpha^\pm)^\dagger = \alpha^\pm
\,.
\label{projector}
\ee
We can use these to show that
\be
\alpha^\pm \psi_\mp = 0 \,\, ; \,\, \alpha^\pm \psi_\pm = \psi_\pm \,\, ;
\,\, \alpha^\pm {\hat\gamma}^0 = {1\over 2} {\hat\gamma}^0 \alpha^\pm 
\,\, ; \,\, \alpha^\pm \gamma_\perp = \gamma_\perp \alpha^\pm \, .
\ee
We will make liberal use of these identities in deriving the light cone 
Hamiltonian in lecture 2.

A particular property of light cone quantization is that it is 
the two component spinor $\psi_+$ above 
that is the dynamical spinor in the light cone 
QCD Hamiltonian $P_{QCD}^-$. Interestingly, the same feature is observed for
fermion fields which obey equal time commutation relations when they are 
boosted to the infinite momentum frame. The dynamical spinors $\psi_+$ are
defined in terms of creation and annihilation operators as 
\be
\psi_+ = \int_{k^+>0} {d^3 k\over{2^{1/4}(2\pi)^3}} \sum_{s=\pm {1\over
2}} \left[ e^{ik\cdot x} b_s (k; x^+) + e^{-ik\cdot x}
d_s^\dagger (k; x^+)
\right] \, ,
\label{normalize}
\ee
where $b_s (k)$ is a quark destruction operator and destroys a quark
with momentum $k$ while $d_s^\dagger (k)$ is an anti--quark creation 
operator and creates an anti--quark with momentum $k$. They obey the 
equal light cone time ($x^+$) anti--commutation relations 
\be
\{b_s (\vec{k},x^+), b_{s^\prime}^\dagger (\vec{k^\prime},x^+)\} =
\{d_s (\vec{k},x^+), d_{s^\prime}^\dagger (\vec{k^\prime},x^+)\} =
(2\pi)^3 \delta^{(3)} (\vec{k}-\vec{k^\prime}) \delta_{s s^\prime} \, .
\ee 

The above definitions ensure that the fermionic contribution to the 
light cone QCD Hamiltonian can be
written as the sum of kinetic and potential pieces,  
$P_{f,QCD}^- = P_{f,0}^- + V_{QCD}$, 
where the kinetic piece of the Hamiltonian is defined as
\be
P_{f,0}^- = \int {d^3 k \over {(2\pi)^3}} \sum_{s = \pm {1\over 2}} 
{(k_t^2 + M^2)\over {2k^+}}\,\,\left(b_s^\dagger (k) b_s (k) + d_s^\dagger 
(k) d_s (k)\right) \, .
\label{fermke}
\ee
These points will become clearer when we explicitly derive the QCD light
cone Hamiltonian in lecture 2.

The gauge field $A^\mu$ has two dynamical components $A_i^a (x)$ with $i=1,2$
in  light cone gauge $A^+=0$. These are defined in terms of
creation--annihilation operators as
\be
A_i^a (x) = \int_{k^+>0} {d^3 k\over{\sqrt{2|k^+|}(2\pi)^3}} \sum_{\lambda =
1,2}\delta_{\lambda i}\left[ e^{ik\cdot x} a_{\lambda}^a (k; x^+) + 
e^{-ik\cdot x}{a_{\lambda}^a}^\dagger (k; x^+)\right] \, , 
\label{gnormalize}
\ee
where the $\lambda$'s here correspond to the two independent polarizations
and ${a_\lambda^a}\dagger (a_\lambda^a)$ creates (destroys) a gluon with 
momentum $k$. They obey the commutation relations
\be
[a_{\lambda}^a (\vec{k}), {a_{\lambda^\prime}^b}^\dagger 
(\vec{k^\prime})] =
(2\pi)^3 \delta^{(3)} (\vec{k}-\vec{k^\prime}) \delta_{ab} \delta_{\lambda 
\lambda^\prime} \, .
\ee
In an analogous fashion to Eq.~\ref{fermke}, the bosonic kinetic energy 
can be written (after normal ordering) as
\be
\int {d^3 k \over {(2\pi)^3}} \sum_{\lambda = 1,2} 
{(k_t^2 + M^2)\over {2k^+}}\,\,a_\lambda^\dagger (k) a_\lambda (k) \, .
\label{boseke}
\ee

We will now discuss the structure of the Poincar\'{e} group on the light cone.
For the field ${\hat \phi}_r$, which here denotes vector or scalar bosons, 
we can define the stress--energy tensor
\be
{\hat T}^{\lambda\nu} = {\hat \Pi}_r^\lambda \partial^\nu {\hat \phi}_r
- {\hat g}^{\lambda\nu} {\cal L} \, ,
\ee
where ${\cal L}$ is the Lagrangean density, and 
${\hat \Pi}_r^\lambda$ is the generalized momentum
\be
{\hat \Pi}_r^\lambda = {\delta {\cal L} \over {\delta (\partial_\lambda 
{\hat \phi}_r)}} \, .
\label{genmom}
\ee
Keep in mind that the carets denote quantities in the usual spacetime
co--ordinates. Define now the following generalized quantity
\begin{equation}
{\hat \Sigma}_{\alpha\beta}^{\mu\nu} = \left\{ \begin {array}{cc}
{1\over 4}[{\hat \gamma}^\mu,{\hat \gamma}^\nu]^{\alpha\beta} &
\mbox{for spinors}\\
({\hat g}_\alpha^\mu {\hat g}_\beta^\nu -{\hat g}_\beta^\mu {\hat
  g}_\alpha^\nu) & \mbox{for vectors} \end{array} \right.
\end{equation} 
One can then define the boost--angular momentum stress tensor 
\be
{\hat J}^{\lambda\mu\nu} = {\hat x}^\mu {\hat T}^{\lambda\nu}
-{\hat x}^\nu {\hat T}^{\lambda\mu} + {\hat \Pi}_r^\lambda {\hat
  \Sigma}_{rs}^{\mu\nu} {\hat \phi}_s \, .
\ee

There are ten conserved currents
\be
\partial_\lambda {\hat T}^{\lambda\mu} &=& 0 \, , \nonumber \\
\partial_\lambda {\hat J}^{\lambda\mu\nu} &=& 0 \, ,
\ee
and correspondingly, ten conserved charges,
\be
{\hat P}^\mu &=& \int d{\hat x}^1\,d{\hat x}^2 \,d{\hat x}^3\,\, {\hat
  T}^{0\mu} \, , \nonumber \\
{\hat M}^{\mu\nu} &=& \int d{\hat x}^1\,d{\hat x}^2 \,d{\hat x}^3 
\left( {\hat x}^\mu {\hat T}^{0\nu}
-{\hat x}^\nu {\hat T}^{0\mu} + {\hat \Pi}_r^0 {\hat
  \Sigma}_{rs}^{\mu\nu} {\hat \phi}_s \right) \, .
\ee
The four components of the energy--momentum vector ${\hat P}^\mu$ and the 
six components of the boost--angular momentum ${\hat M}^{\mu\nu}$ comprise 
the ten generators of the Poincar\'{e} group~\footnote{The Poincar\'{e} 
group is a sub--group of the conformal group which contains 15 generators, 
the additional generators being 4 conformal 
transformations and 1 dilatation.}. These generators satisfy the Poincar\'{e} 
algebra
\be
[{\hat P}^\mu,{\hat P}^\nu] &=& 0 \,\,;\,\, [{\hat M}^{\mu\nu},{\hat
  P}^\rho] = i \left({\hat g}^{\nu\rho} {\hat P}^\mu- 
{\hat g}^{\mu\rho} {\hat P}^\nu \right) \, , \nonumber \\
\left[{\hat M}^{\mu\nu},{\hat M}^{\rho\sigma}\right] &=& i\left(
{\hat g}^{\mu\sigma} {\hat M}^{\nu\rho}+{\hat g}^{\nu\rho} 
{\hat M}^{\mu\sigma}-{\hat g}^{\mu\rho} {\hat M}^{\nu\sigma}
-{\hat g}^{\nu\sigma} {\hat M}^{\mu\rho}\right) \, .
\label{carcomm}
\ee
The six components of the boost--angular momentum can be further split into
the three generators of rotations ${\hat M}^{ij} = \varepsilon^{ijk}
{\hat J}^k$ (where ${\hat J}^k$ is the angular momentum operator) and
three generators of boosts ${\hat K}^i = {\hat M}^{i0}$. In the 
language of Ref.~\cite{Dirac}, these are referred to as kinematic and 
dynamic operators, respectively, since the former is independent of the 
interaction while the latter isn't.

Transforming the above to light cone co--ordinates, we obtain 
$P^\mu \equiv (P^+,P^1,P^2,P^-)$, where $P^\pm = ({\hat P}^0\pm {\hat
  P}^3)/\sqrt{2}$, and
\begin{equation}
M^{\mu\nu} = \left( \begin{array}{cccc} 
0 & -S_1 & -S_2 & K_3 \\
S_1 & 0 & J_3 & B_1 \\
S_2 & -J_3 & 0 & B_2 \\
-K_3 & -B_1 & -B_2 & 0 \\ 
\end{array} \right) \, .
\end{equation}
Above we used the following definitions
\be
B_1 &=& {(K_1 + J_2)\over {\sqrt{2}}} \,\, ; \,\, B_2 = {(K_2 - J_1)
\over {\sqrt{2}}} \, , \nonumber \\
S_1 &=& {(K_1 - J_2)\over {\sqrt{2}}} \,\, ; \,\, S_2 = {(K_2 + J_1)
\over {\sqrt{2}}} \, . \nonumber
\ee
The commutation relations among the $M^{\mu\nu}$'s and the $P^\mu$'s are 
of course the same as in Eq.~\ref{carcomm}. The operators $B_1$ and $B_2$ 
are kinematic and boost the system in the x and y directions respectively. 
In addition, the operators $J_3$ and interestingly, $K_3$ are kinematic 
and rotate the system in the x--y plane and boost it in the longitudinal
direction respectively.

An interesting observation by Susskind~\cite{Susskind} related to the above 
is that the commutation relations among the seven generators $P^\pm, 
{\vec{P}}_t,J_3, B_1$ and $B_2$ are the same as the commutation relations 
among the symmetry operators of non--relativistic quantum mechanics in
two dimensions. Indeed, one can formally make the correspondence,
\begin{itemize}
\item
$ P^- \longrightarrow$  Hamiltonian.

\item
${\vec{P}}_t \longrightarrow$ Momenta. 

\item
$P^+ \longrightarrow$  Mass.

\item
$J_3 \longrightarrow$ Angular Momentum.

\item
${\vec{B}}_t \longrightarrow$ generators of Galilean boosts in x--y
plane.
\end{itemize}

These seven generators obey the commutation relations
\be
[P^-,P_t] = [P^-,P^+] = [P_t,P^+]=[J_3,P^-]=[J_3,P^+]=[B_t,P^+]=0 \, .
\ee
and
\be
[J_3, P^t] = i\epsilon_{tl}P^l\,, [J_3,B^t]=i\epsilon_{tl}B^l\,,
[B^t,P^-] = -iP^t\,, [B^t,P^l]=-i\delta_{tl}P^+ \, .   
\ee
Above $\epsilon_{ij}$ is the Levi--Civita tensor in two dimensions. 
Since they are kinematic operators, they leave the planes of $x^+ = constant$
invariant under their operations.

Susskind, Bardacki \& Halpern~\cite{BarHal} and Kogut \& Soper have shown
that the above mentioned isomorphism is responsible for the non--relativistic
quantum--mechanical structure of quantum field theories on the light cone.
The simplest illustration of this isomorphism is the fact that the free
particle Hamiltonian takes the form
\be
H\equiv P^- = {P_t^2 + M^2 \over {2P^+}} \, . \nonumber
\ee
Recalling the form of the energy in two dimensional quantum mechanics, we
obtain the isomorphisms above. For QED and QCD, the
above form is modified by the addition of a potential term which we will
discuss in detail in lecture 2.
Finally, we should mention that the other kinematic operator, $K_3$, the
boost operator in the longitudinal direction, serves to rescale the other 
operators
\be
\exp(i\omega K_3) P^- \exp(-i\omega K_3) &=& \exp(\omega) P^- \, . \nonumber \\
\exp(i\omega K_3) J_3 \exp(-i\omega K_3) &=&  J_3 \, . \nonumber \\
\exp(i\omega K_3)\vec{S} \exp(-i\omega K_3) &=& \exp(-\omega) \vec{S} \, .
\ee
This property of $K_3$ will come in handy in lecture 3.

\newpage

\section*{ Lecture 2: The light cone QCD Hamiltonian.}
\vskip 0.15in

In this lecture, we will derive an explicit form for the light cone 
QCD Hamiltonian making use of the light cone constraint relations.
Consider first the fermionic part of the QCD action
\be
S_F = \int d^4 x \, {\psibar} \left(P\!\dirsh + M\right) \psi \, . \nonumber
\ee
Above, $P^\mu = -iD^\mu\equiv -i(\partial^\mu-igA^\mu)$. For convenience,
we will not write the integral $\int d^4 x$, in the following but it must 
be understood to be there.
Then writing out the above action explicitly, 
\be
S_F = {\psibar}\gamma^-{P_-}\psi+{\psibar}\gamma^+{P_+}\psi +
{\psibar}\gamma^t{P_t}\psi +
{\psibar}M\psi \, .\nonumber 
\ee
Consider now the first term in the above:
\be
{\psibar} \gamma^- P_\psi = \psi^\dagger {\hat \gamma}^0\gamma^- P_-\psi
\rightarrow \sqrt{2}\psi_{-}^\dagger P_-\psi_- \, .
\ee 
To dissect the above, we first decomposed ${\hat \gamma}^0 = (\gamma^+ + 
\gamma^-)/\sqrt{2}$, made use of $(\gamma^-)^2 =0$, and the properties of
the projector $\alpha^+$ in Eq.~\ref{projector} to obtain the RHS. 
Similarly, it is recommended to the serious student that he or she show that 
\be
{\psibar}\gamma^+{P_+}\psi &=& \sqrt{2}\psi_{+}^\dagger P_+\psi_{+} 
\, ,\nonumber \\
{\psibar}\gamma^t {P_t}\psi &=& {1\over \sqrt{2}}\left(\psi_{-}^\dagger\gamma^+
P_t\!\!\!\!/ + \psi_{+}^\dagger\gamma^- P_t\!\!\!\!/\psi_-\right) \, , 
\nonumber\\
{\psibar}M\psi &=& M\left(\psi_{+}^\dagger{\hat \gamma}^0\psi_{-}+
\psi_{-}^\dagger{\hat \gamma}^0\psi_{+}\right) \, .
\ee
One then obtains
\be
S_F &=& \sqrt{2} \psi_{-}^\dagger P_-\psi_{-} + 
\sqrt{2}\psi_{+}^\dagger P_+\psi_{+} 
+{1\over \sqrt{2}}\left(\psi_{-}^\dagger\gamma^+
P_t\!\!\!\!/ + \psi_{+}^\dagger\gamma^- P_t\!\!\!\!/\psi_-\right)\nonumber\\
&+&M\left(\psi_{+}^\dagger{\hat \gamma}^0\psi_{-}+
\psi_{-}^\dagger{\hat \gamma}^0\psi_{+}\right) \, ,
\label{oldferm}
\ee  
where we have written the fermionic piece of the action in terms of the 
two--spinors $\psi_{-}$ and $\psi_{+}$ and their hermitean conjugates.

Following Eq.~\ref{genmom}, the momenta conjugate to these two--spinor 
fields are
\be
{\Pi_+} &=& 
{\delta {\cal L}\over {\delta (\partial_+\psi_+)}} = \left({\sqrt{2}
\over i}\right)\psi_{+}^\dagger \, , \nonumber\\
\Pi_- &=& {\delta {\cal L}\over {\delta (\partial_+\psi_-)}} =0 \, .
\ee
Since trivially $[\Pi_-,\psi_-]=0$, the two--spinor $\psi_-$ is, unlike 
$\psi_+$, not an independent quantum field. We will now show that one may 
derive a constraint equation (i.e., independent of the light cone time $x^+$)
for $\psi_-$ in terms of the dynamical field $\psi_{+}$. The light cone constraint relations can be obtained from the operator equations of motion. In this 
case, it is the Dirac equation $(P\!\!\!\!/+M)=0$, or
\be
(-i\partial_- -gA_-)\gamma^-\psi + (-i\partial_+ - gA_+)\gamma^+\psi
+ \sum_{j=1}^2 (-i\partial_j -gA_j)\gamma^j\psi + M\psi =0 \, .
\ee
Multiply the above through by $\gamma^+$. Since $(\gamma^+)^2=0$, this 
projects out the $x^+$--light cone time--dependence in the above and we 
obtain (after liberally using our projection operator tricks from 
Eq.~\ref{projector}) the equation
\be
\sqrt{2}P_- \psi_- = -{\hat \gamma}^0 (P_t\!\!\!\!/+M)\psi_+ \, .
\ee
In light cone gauge, $A_- = -A^+ =0$, hence $P_-=(-i\partial_- -gA_-)
\rightarrow -i\partial_-$. With this gauge condition therefore, one can 
easily invert the $P_-= -P^+$ operator and one obtains the light cone 
constraint equation
\be
\psi_- = {{\hat \gamma}^0\over {\sqrt{2}P^+}}\left(P_t\!\!\!\!/+M\right)\psi_+ 
\, . 
\label{fermconst}
\ee
Thus for light cone time $x^+$, $\psi_-$ is determined completely by 
$\psi_+$ at that time. Only the two components of the spinor $\psi$ corresponding to $\psi_+$ are independent dynamical fields on the light cone.

We can now use the above obtained constraint equation to replace $\psi_-$ 
in Eq.~\ref{oldferm} for $S_F$. For instance, 
\be
\sqrt{2}\psi_{-}^\dagger P_{-}\psi_{-} &=& \sqrt{2} \left({{\hat\gamma}^0 
\over 
{\sqrt{2}P^+}}({P_t}\!\!\!\!/ + M)\psi_{+}\right)^\dagger P_{-}
\left({{\hat\gamma}^0 \over{\sqrt{2}P^+}}({P_t}\!\!\!\!/ + M)\psi_{+}\right)
\nonumber \\
&\longrightarrow& {(-)\over\sqrt{2}}\psi_{+}^\dagger (M-{P_t}\!\dirsh ) 
{1\over P^+} ({P_t}\!\dirsh +M)\psi_{+} \, . \nonumber
\ee
Now rescale the fields $\psi\rightarrow 2^{-1/4}\psi$~\footnote{This
rescaling gets rid of the $\sqrt{2}$ factors in the action. This also
explains the peculiar $2^{-1/4}$ normalization factor in
Eq.~\ref{normalize} for the properly normalized $\psi_+$ field.}.  As an
exercise, the reader should use the light cone constraint equation
above, the properties of the projection operators $\alpha^\pm$ in
Eq.~\ref{projector} and those of the light cone gamma matrices to
first substitute for $\psi_-$ everywhere and then 
demonstrate the following identities,
\be 
{M\over \sqrt{2}}\left(\psi_{+}^\dagger {\hat\gamma}^0\psi_{-} + 
\psi_{-}^\dagger {\hat\gamma}^0\psi_{+}\right) &=& {M\over 2}\left(
\psi_{+}^\dagger \gamma^- \psi_{-} + \psi_{-}^\dagger \gamma^+\psi_{+}\right)
\nonumber \\ 
{1\over 2} \left(\psi_{-}^\dagger \gamma^+ ({P_t}\!\dirsh +M)\psi_+ +
\psi_{+}^\dagger \gamma^- ({P_t}\!\dirsh +M)\psi_- \right) &=& \psi_{+}^\dagger 
(M-{P_t}\!\dirsh ){1\over P^+} ({P_t}\!\dirsh +M)\psi_+ \, . \nonumber
\ee
Putting these together with the other term above, our result for the fermionic
action expressed solely in terms of the dynamical two--spinor $\psi_+$ is
\be
S_F = -\psi_{+}^\dagger P^- \psi_+ + {1\over 2}\, \psi_{+}^\dagger 
(M-{P_t}\!\dirsh ){1\over P^+}({P_t}\!\dirsh +M)\psi_+ \, .
\ee

We now turn to the bosonic contribution to the action,
\be
S_B = {1\over 4} F_{\munu}^a F^{\munu,a} \, ,
\ee
and following a procedure analogous to the fermionic case, shall write it
in terms of $A_t$, the two transverse, dynamical components of the gauge 
field $A^\mu$. We have seen earlier that the choice of light cone gauge
$A_-=-A^+=0$ greatly simplifies the light cone constraint relation for the 
fermions. In this gauge, the various components of the field strength 
tensor also simplify to
\be
F_{+-}^a &=& -\partial_{-} A_+^a \, , \nonumber\\
F_{t+}^a &=& \partial_t A_{+}^a - \partial_+ A_{t}^a + gf^{abc} A_t^b A_+^c
\, , \nonumber\\
F_{t-}^a &=& -\partial_- A_t^a \equiv -E_t^a \, . \nonumber
\ee
In addition, there are of course the purely transverse pieces $F_{ij}$; 
with $i,j=1,2$. 
From the above it is evident that there is no (light cone) time 
derivative $\partial_+ A_+$ in the action. The field $A_+$ therefore has
no momentum conjugate and we may use the operator equations of motion to
eliminate the field $A_+=-A^-$.

The equations of motion are the Yang--Mills equations of course. The 
light cone constraint equation is just Gauss' law on the light cone since 
it must be valid at all times. This condition is then
\be
(D_t F^{t+})^a + (D_- F^{-+})^a = -J^{+,a} \Longrightarrow -{\partial_-^2}
A^{-,a} = J^{+,a} + (D_t E_t)^a \, ,\nonumber
\ee
where $E_t^a$ are the two transverse components of the electric field and
$D_t$ is the covariant derivative $\partial_t -igA_t$.
We can write our light cone constraint equation for $A^-$ compactly below
as
\be 
A^{-,a} = {1\over {(P^+)^2}} \left(J^{+,a} + (D_t E_t)^a\right) \, .
\label{boseconst}
\ee
Returning to the action
\be
S_B = {1\over 4} F^2 = {1\over 4}F_t^2 -F_{t+}^a F_{t-}^a -{1\over 2} 
F_{+-}^a F_{+-}^a \, , \nonumber
\ee
substituting the expressions for the field strength components in terms of
the gauge fields and performing an integration by parts, we obtain
\be
S_B = {1\over 4}F_t^2 -A_+^a (D_t E_t)^a -{1\over 2}(\partial_- A_+)^2 
-(\partial_- A_t^a)\,(\partial_+ A_t^a) \, .
\ee

Before we substitute for $A_+$ above, we will first write out the full
action $S_{QCD} = S_F + S_B -J_{ext}\cdot A$:
\be
S_{QCD} &=& -\psi_+^\dagger (-i\partial^- -gA^-)\psi_+ + {1\over 2} 
\psi_+^\dagger (M-{P_t}\!\dirsh ){1\over P^+}({P_t}\!\dirsh +M)\psi_+
+ {1\over 4} F_t^2 \nonumber \\
&-& A_+^a (D_t E_t)^a 
-{1\over 2}(\partial_- A_+)^2 
-(\partial_- A_t^a)\,(\partial_+ A_t^a) - J_{ext}^+ A_+ \, .\nonumber
\ee
Consider the first term above. We can write this as
\be
-\psi_{+}^\dagger \left(-i\partial^- -gA^-\right)\psi_+ = -i\psi_+^\dagger 
\partial_+ \psi_+ - J_{dyn}^+ A_+ \, , \nonumber
\ee
where $J_{dyn}^{+,a} = {\psi_+^a}^\dagger \lambda^a \psi_+$. (The $\lambda^a$ 
are the Gell--Mann SU(3) matrices.) We now substitute the above result in our
expression for the action and after 

a) defining 
$J^+ = J_{dyn}^+ + J_{ext}^+$\,, 

b) performing an integration by parts, 

c) making use of the constraint relation Eq.~\ref{boseconst} to eliminate 
$A_+$, 

we obtain finally,
\be
S_{QCD} &=& -i\psi_+^\dagger \partial_+ \psi_+ - (\partial_- A_t)\,(\partial_+
A_t) + {1\over 4} F_t^2 
+ {1\over 2} \psi_+^\dagger (M-{P_t}\dirsh ){1\over P^+}({P_t}\dirsh +M)
\psi_+ \nonumber \\
&+& {1\over 2} (J^+ + D_t E_t){1\over {(P^+)^2}}(J^+ + D_t E_t) \, .
\ee

The final step before we obtain the Hamiltonian is to identify the momenta
conjugate to the dynamical fields (now with the proper normalization!),
\be
\Pi_+^{fermi} &=& {\delta S_{QCD} \over{\delta (\partial_+\psi_+)}} = 
-i \psi_+^\dagger \, . \nonumber \\
\Pi_t^{bose} &=& {\delta S_{QCD} \over {\delta (\partial_+ A_t)}} = 
-\partial_-A_t \equiv -E_t \, .
\ee
Writing out the fields and their momentum conjugates in terms of the 
creation and annihilation operators introduced in Eqs.~\ref{normalize} and
~\ref{gnormalize}, and making use of their commutation relations, the reader
may confirm that
\be
[\Pi_t^{bose}(x), A_t(x^\prime)] = 
\{\Pi_+^{fermi}(x), \psi_+(x^\prime)\} = i\delta^{(3)} (x-x^\prime)\, .
\ee
The Hamiltonian density in our convention is defined as
\be
H\equiv P_{QCD}^- = S_{QCD} - \Pi_{+}^{fermi}\partial_+ \psi_+ -
\Pi_{t}^{bose}  \partial_+ A_t \, , \nonumber
\ee
We can therefore  write our final expression for the Hamiltonian density as 
\be
P_{QCD}^- &=& {1\over 4}F_t^2 + {1\over 2} (J^+ + D_t E_t){1\over {(P^+)^2}}
(J^+ + D_t E_t) \nonumber \\
&+& {1\over 2} \psi_+^\dagger (M-{P_t}\!\dirsh ){1\over P^+} 
({P_t}\!\dirsh +M)\psi_+ \, .
\label{QCDham}
\ee

We have therefore succeeded in obtaining the light cone
Hamiltonian in QCD, expressed solely in terms of the two--spinor $\psi_+$ 
and $A_t$, the two transverse components of the gauge field. The following
observations can be made regarding the above expression. Firstly, one can show 
straightforwardly that the light cone Hamiltonian can be written as
\be
P_{QCD}^- = P_0^- + V_{QCD} \, ,
\label{potential}
\ee
where $P_0$ (the sum of the RHS of Eqs.~\ref{fermke} and~\ref{boseke}) 
is the piece of the Hamiltonian not containing any factors of the coupling 
$g$ and $V_{QCD}$ is the rest, which can also be written out in terms of 
creation--annihilation operators. Furthermore, the ground state of the 
non--interacting Hamiltonian $P_0^-$ is also, remarkably, the ground state of 
the full Hamiltonian. This is the meaning behind statements one may have heard
that the light cone vacuum is `trivial'. Because the vacuum is trivial, 
one may simply construct 
any eigenstate of the full Hamiltonian in terms of a complete Fock eigen--basis
corresponding to eigenstates of the non--interacting Hamiltonian. As we shall 
demonstrate in the next lecture with a specific example, this point forms the 
basis for the quark--parton model in quantum field theory.

Just as in non--relativistic quantum mechanics then, one can use
light cone time ordered perturbation theory to construct these
states. Unfortunately, there is no room to discuss time ordered
perturbation theory here but it will be discussed in the ``long''
version of these lectures~\cite{Rajlater}. 

There is one point we have not mentioned thus far but it threatens the entire
pretty picture above. This has to do with the terms $1/P^+$ and $1/(P^+)^2$
above. Recall that they were obtained by inverting the light cone constraint
equations in light cone gauge. Clearly, that operation and these terms are
not well defined for $P^+ =0$.  The simple vacuum is thus only deceptively so
and all the complications are now hidden in the zero--mode. That this would
be the case should have been clearer in retrospect. Defining the operator
$1/P^+$ requires knowing the boundary conditions of the fields at large
distances and therefore, should be sensitive to confining and chiral symmetry
breaking effects. Attempts to regulate the zero mode, a well know example of
which is discretized light cone quantization~\cite{BPP}, also result in a
non--trivial vacuum.  On the other hand, perturbative physics should not be
terribly sensitive to how fields are regulated at large distances. Different
`epsilon' prescriptions corresponding to different boundary conditions at
infinity give the same short distance physics~\cite{Hoyer}. 
The justification of
the above approach is therefore the success of the parton model in describing
physics at large transverse momenta in QCD. The program to describe
non--perturbative physics in the same framework is very advanced and we refer
the reader to Ref.~\cite{BPP} to read of the latest developments.

\newpage

\section*{Lecture 3: High energy Eikonal scattering and the parton 
model in QED.}
\vskip 0.15in

In the previous lectures we developed some of the basic formalism of
light cone field theory. We will now apply this formalism to a
specific example; high energy scattering of an electron from an
external potential in QED.  We will show how one recovers the standard
Eikonal picture in this formalism. More importantly, our results
clearly can be interpreted in terms of a parton model picture of high
energy scattering. This lecture closely follows the excellent paper of
Bjorken, Kogut and Soper~\cite{BjKogutSoper} where this example and
others are discussed. For convenience, we will also use their ``-2''
convention (for eg., $A_- = A^+$ and $A_t = -A^t$).

The light cone Hamiltonian in QED is similar to the QCD Hamiltonian 
derived above in Eq.~\ref{QCDham} and of course much simpler.  
To treat the problem of
scattering off an external potential, we introduce an external potential 
${\bf a}_\mu$ using the gauge invariant minimal substitution 
$p_\mu\rightarrow p_\mu -g{\bf a}_\mu$. The QED Hamiltonian including the 
external potential ${\bf a}_\mu$ is then
\be
P_{scatt}^- (x^+) &=& \int d^2 x_t\, dx^- \Bigg\{ e{\bf a}_+ 
\psi_+^\dagger \psi_+
+ {1\over 2} e^2 \psi_+^\dagger \psi_+ {1\over {(p^+ -e{\bf a}^+})^2}
\psi_+^\dagger
\psi_+ \nonumber \\ 
&+& 
\psi_+^\dagger \big(M-i\vec{\sigma}\cdot (\vec{p}-e\vec{A}-e\vec{{\bf a}})
\big)
{1\over {2(p^+ -{\bf a}^+)}}\big(M + i\vec{\sigma}\cdot (\vec{p}-e\vec{A}-
e\vec{{\bf a}})\big)\psi_+ \nonumber \\
&+& e\psi_+^\dagger\psi_+ {1\over {p^+ - e{\bf a}^+}}\vec{p}\cdot \vec{A}
+ {1\over 2}\sum_{t=1,2} A^t {\vec{p}}^{\,2} A^t\Bigg\}
\, .
\ee
  
Note that one can define
\be
\left[ {1\over {p^+ -{\bf a}^+}} \psi_+\right](x) = \int d\xi\, {1\over {2i}}
\epsilon (x^- -\xi) \exp\left(-ie\int_\xi^{x^-} d\xi^\prime {\bf a}^+(x^+,
x_t,\xi^\prime)\right)\,\psi_+(x^+,x_t,\xi) \, ,\nonumber \\
\ee
where $\epsilon$ is the sign function. This can be checked by multiplying 
through by $p^+ -e{\bf a}^+$~\footnote{In QCD, the sole change is to replace 
the exponential on the RHS by a path ordered exponential.}.    

Now write $P_{scatt}^- = P_{QED}^- + V$, where $P_{QED}$ is the usual
time independent QED Hamiltonian with ${\bf a}^\mu=0$ and $V(x^+) =
P_{scatt}^- - P_{QED}^-$.  We wish to construct the scattering matrix
$S_{fi}$ in the interaction picture. In the usual quantum mechanical
treatment,
\be
\psi_I (x^+,x_t,x^-) &=& e^{iP_{QED}^- x^+} \psi_+ (0,x_t,x^-) e^{-iP_{QED}^-
x^+} \, . \nonumber \\
{\vec{A}}_I (x^+,x_t,x^-) &=& e^{iP_{QED}^- x^+} \vec{A} (0,x_t,x^-) e^{-iP_{QED}^-
x^+} \, .
\ee
Then, the scattering matrix for the scattering of an electron off an external 
potential is given by
\be
S_{fi} = <f|T\left\{ \exp\left(-i\int dx^+ V (x^+)\right)\right\}|i> \, ,
\ee
where `T' denotes light cone time ordering and $|i>$ and $|f>$ are asymptotic 
states which are eigenstates of the QED Hamiltonian $P_{QED}$. They can 
thus be evaluated in Rayleigh--Schr\"{o}dinger perturbation theory (see the
discussion at the end of lecture 2).

We want to compute the scattering matrix in the high energy scattering 
limit $P_i,P_f\rightarrow \infty$. Consider the states $|I>$ and $|F>$, which 
may be states in the rest frame of the electron. They are related by 
boosts to the states $|i>$ and $|f>$ above.  Then $|i> = e^{-i\omega 
K_3}|I>$ and $|f> = e^{-i\omega K_3}|F>$, where $K_3$ (defined previously 
in lecture 1) is the generator of boosts in the longitudinal direction. In
QED, $K_3$ is the operator
\be
K_3 = \int d^2 x_t \,dx^-\, x^- \left[ {i\over 2} \psi_+^\dagger 
\stackrel{\leftrightarrow}{\partial_-} \psi_+ + (\partial_- A_t)\,(\partial_-
A_t)\right]_{x^+=0} \, ,
\label{defk3}
\ee
and $\omega$ is the rapidity corresponding to the boost. 
The scattering matrix element between the scattering states in the rest frame 
is then
\be
<F|e^{i\omega K_3}T\left\{ \exp\left(-i\int dx^+ V(x^+)\right)\right\} 
e^{-i\omega K_3}|I> \, .\nonumber
\ee

Using the definition of path ordered exponentials, this relation can be
written as
\be
<F|T\left\{ \exp\left(-i\int dx^+ e^{i\omega K_3}\, V(x^+)\,e^{-i\omega K_3}\right)
\right\} |I> \, .
\label{restmrx}
\ee
A great advantage of the light cone formalism is that the fields transform 
simply under boosts. We have 
\be
e^{i\omega K_3} \psi_I (x^+,x_t,x^-) e^{-i\omega K_3} &=& e^{\omega/2} 
\psi_I (e^{-\omega}x^+,x_t,e^{\omega}x^-) \, . \nonumber \\
e^{i\omega K_3} A_I (x^+,x_t,x^-) e^{-i\omega K_3} &=&
A_I (e^{-\omega}x^+,x_t,e^{\omega}x^-) \, . \nonumber 
\label{fieldboost}
\ee
The above can be shown explicitly by computing the commutators $[K_3,\psi_I]$
and $[K_3,A_I]$ using the definition of $K_3$ in Eq.~\ref{defk3}. The field
${\bf a}$ however commutes with $K_3$ and therefore does not transform under
boosts.

Consider now the argument of the exponential in Eq.~\ref{restmrx}. We can 
show that all but one of the terms in $V(x^+)$ are invariant under the 
boost operation. For example,
\be
e^{i\omega K_3}\,\psi_I^\dagger \psi_I {1\over {(P^+)^2}}\psi_I^\dagger \psi_I
\,e^{-i\omega K_3} \longrightarrow \psi_I^\dagger \psi_I 
{1\over {(P^+)^2}}\psi_I^\dagger \psi_I  \, .\nonumber
\ee
Above, we have used the fact that elements of the Lorentz group are 
simply rescaled by boosts, 
$e^{i\omega K_3}P^+ e^{-i\omega K_3}=e^{\omega}P^+$, 
as well as Eq.~\ref{fieldboost}. 
The only term that does not remain invariant is
\be
e {\bf a}_+ \psi_I^\dagger \psi_I \stackrel{K_3}{\longrightarrow}e^{\omega} 
e{\bf a} \psi_I^\dagger \psi_I \, . \nonumber 
\ee
Hence, 
\be
e^{i\omega K_3} V(x^+) e^{-i\omega K_3} &=& \int d^2 x_t\, dx^- \,e^{\omega} 
e \,{\bf a}_+ (x^+,x_t,x^-) \psi_I^\dagger (e^{-\omega}x^+,x_t, e^{\omega}x^-)
\psi  (e^{-\omega}x^+,x_t, e^{\omega}x^-) \nonumber \\
&+& O(e^{-\omega})\, .\nonumber
\label{intext}
\ee
Now let $x^-\rightarrow e^{\omega}x^-$ above. Then
\be
e^{i\omega K_3} V(x^+) e^{-i\omega K_3} &=& \int d^2 x_t\, dx^-\, 
e {\bf a}_+ (x^+,x_t,e^{-\omega} x^-) \psi_I^\dagger (e^{-\omega}x^+,x_t, x^-)
\psi  (e^{-\omega}x^+,x_t,x^-) \nonumber \\
&+& O(e^{-\omega})\, .
\ee
Going to the infinite rapidity limit $\omega\rightarrow \infty$ corresponding 
to very high energy scattering, we note from the above that the operators are
all evaluated at $x^+=0$ so the time ordering in $x^+$ is irrelevant in that
limit. Then one can show in that limit (and this is a subtle point) that
\be
S_{fi} = <F|{\cal P}|I> + O(e^{-\omega}) \equiv <f|{\cal P}|i> + O(e^{-\omega}) \, .
\label{scatmat}
\ee
Thus  we have expressed $S_{fi}$ again in terms of the 
states $|i>,|f>$, thereby 
demonstrating the Lorentz invariance of these states in the infinite momentum 
limit. Also, above
\be
{\cal P} = \exp\left( -i \int d^2 x_t\, \chi(x_t)\rho(x_t) \right) \, ,
\ee
where 
\be
\chi(x_t) = e\int dx^+ {\bf a}_+ (x^+, x_t,0)\, ,
\ee
and
\be
\rho(x_t) = \int dx^- \psi_I^\dagger (0,x_t,x^-)\psi_I (0,x_t,x^-) \, .
\ee
We have therefore recovered the well known eikonal scattering limit in
QED.

We shall now show that the above derivation has a deep connection with
the parton model. The asymptotic `in' state of the electron, $|i>$, is 
an eigenstate of the QED Hamiltonian $P_{QED}^-$. We can expand $|i>$ in
terms of the ``bare'' quanta~\footnote{These are eigenstates (in QED!) of 
$P_0^-$ in 
Eq.~\ref{potential}.} associated with the fields $\psi_+(0,x_t,x^-)$
and $A_t(0,x_t,x^-)$ at $x^+=0$: 
\be
& &|i> = \int d^2 k_t \int_{k^+>0} {dk^+\over k^+} \sum_\lambda \Bigg\{
g(k_t,k^+,\lambda) a^\dagger (k_t,k^+,\lambda)|0> \nonumber \\
&+& \int d^2 k_{t1} {dk_1^+\over {k_1^+}} \int d^2 k_{t2} {dk_2^+\over {k_2^+}}
\sum_{s_1,s_2} h(\vec{k_1},\vec{k_2}; s_1,s_2)b^\dagger(\vec{k_1};s_1)
d^\dagger (\vec{k_2};s_2)|0> + \cdots \Bigg\} \,\, .
\ee
The creation and annihilation operators introduced here are the same as 
those in lecture 1. The coefficient $h$ above can be interpreted simply as the
amplitude for $|i>$ to contain a {\it bare electron} with momenta 
$\vec{k_1}$ and spin $s_1$, and a {\it bare positron} with momenta $\vec{k_2}$ 
and spin $s_2$. It was shown first by Drell, Levy and Yan that the amplitude 
squared for an arbitrary number of parton eigenstates, integrated over 
phase space could be simply related to the structure functions 
$W_1,W_2$~\cite{DrellYanLevy}.

We can also see this here if we similarly 
expand $|f>$ in terms of the bare quanta. The scattering matrix $S_{fi}$ 
in Eq.~\ref{scatmat} can then be evaluated if we move ${\cal P}$ past the
creation--annihilation operators till it acts on $|0>$:
\be
{\cal P}b^\dagger d^\dagger a^\dagger\cdots a^\dagger|0> =
{\cal P}b^\dagger {\cal P}^{-1}\cdots {\cal P}a^\dagger {\cal P}^{-1}
{\cal P}|0> \, .
\ee
Since it is evident that ${\cal P}$ is invariant under translations in the
$x^-$ direction, it commutes with the generator of $x^-$ translations--$P^+$.
One can check that ${\cal P}|0> = |0>$. This follows formally by expanding 
${\cal P}$ and requiring that the operators in $\rho(x_t)$ are normal ordered.
How do the creation--annihilation  operators transform with ${\cal P}$?
Using the light cone commutation relations, $\{\psi_+(x),\psi_+^\dagger(x^\prime)\} = \delta^{(3)}(x-x^\prime)$, we find
\be
{\cal P}\psi_+^\dagger(0,x_t,x^-){\cal P}^{-1} = \exp\left(-i\chi\right) 
\psi_+^\dagger(0,x_t,x^-) \, .
\ee
Fourier transforming the above, and using Eq.~\ref{normalize}, we obtain
for the electron creation operator
\be
{\cal P}b^\dagger (k_t,k^+,s) {\cal P}^{-1} = \int {d^2 k_t^\prime \over 
{(2\pi)^2}} b^\dagger(k_t^\prime,k^+,s) {\tilde {\cal P}}(k_t^\prime-k_t)
\, ,
\ee
where (with $q_t=k_t^\prime - k_t$)
\be
{\tilde {\cal P}}(q_t) = \int d^2 x_t e^{-i\vec{q}_t\cdot \vec{x}_t}
e^{-i\chi(x_t)} \, .
\ee
Similarly for the positron creation operator
\be
{\cal P}d^\dagger (k_t,k^+,s) {\cal P}^{-1} = \int {d^2 k_t^\prime \over 
{(2\pi)^2}} d^\dagger(k_t^\prime,k^+,s) {\tilde {\cal P}_c}(k_t^\prime-k_t)
\, ,
\ee
with
\be
{\tilde {\cal P}_c}(q_t) = \int d^2 x_t e^{-i\vec{q}_t\cdot \vec{x}_t}
e^{+i\chi(x_t)} \, .
\ee
Finally, ${\cal P}a^\dagger {\cal P}^{-1}$, since all the operators in 
${\cal P}$ commute with $a^\dagger$.

What we have learnt from the above is that when a high energy {\it bare 
electron} or {\it bare positron} interacts with a potential at $x_t$, the
net effect is to multiply its wavefunction by the eikonal phase $e^{-i\chi}$
or $e^{+i\chi}$ respectively. The following physical picture then emerges
from our manipulations above.
\begin{itemize}

\item
The scattering of high energy particles (denoted here by `$|i>$', which is
an eigenstate of the Hamiltonian $P_{QED}^-$) is 
not simple--i.e., it cannot be described by a simple overall phase.

\item
However, due to the ``potential'' structure of QED on the light cone, the 
{\it physical} particle states ($|i>$) can be expanded in a complete basis of
multi--parton eigenstates (eigenstates of $P_{0,QED}^-$).

\item
The scattering of these partons is simple--they acquire an eikonal phase in
the scattering.

\item
The mutual interactions of partons in the physical state $|i>$ is complex, 
but as the rapidity $\omega\rightarrow \infty$, these interactions 
are slowed down by time dilation. Recall that in Eq.~\ref{intext}, the only
term that survives is the one that contains the coupling to the external 
field ${\bf a}$ and all the other terms which contain the interactions of
the partons with each other are suppressed. 
\end{itemize}

Chronologically, one can view the scattering as follows. Partons in the 
initial state interact strongly for $-\infty<x^+<0$ with the potential 
$V_{QED}$. At $x^+=0$, each individual parton scatters simply off the external 
potential, acquiring an eikonal phase. For $0<x^+<\infty$, the partons then
again interact among each other with the potential $V_{QED}$. This picture
of scattering is also known as the impulse approximation. It explains the
striking phenomenon of Bjorken scaling observed in deep inelastic scattering
at very large momentum transfers.

Finally, for completeness, we will mention that the cross--section for
electron scattering off an external potential is given by
\be
d\sigma = \int_{k_1^+,\cdots , k_n^+ >0}
 {d^2 k_{t1} dk_1^+\over {(2\pi)^3 k_1^+}}\cdots {d^2 k_{tn}
dk_{n}^+\over {(2\pi)^3 k_{n}^+}} (2\pi)\delta(k^+ - \sum_{i=1}^n k_i^+)
\times |<f|{\cal T}|i>|^2 \, ,
\ee
where the transition amplitude is defined as
\be
<f|{\cal T}|i> = <f|U(\infty,0)[{\cal P}-1] U(0,-\infty)|i> \, .
\ee
Above, $U$ is the light cone analog of the usual unitary evolution operator 
in quantum mechanics.

\newpage

\section*{Lecture 4: Bjorken scaling and light cone Fock 
distributions.}
\vskip 0.15in

In this last lecture, we will discuss the ``light cone'' limit $x^2\rightarrow 
 0$ of deep inelastic scattering, in QCD. For very large momentum 
transfers, in this limit, one observes the phenomenon known as 
Bjorken scaling. 
Unfortunately, we will not have room for a discussion of the renormalization 
group ideas which predict, in QCD, the experimentally observed logarithmic 
violations of Bjorken scaling. These will be presented in the ``longer'' 
version of these lectures at a later date~\cite{Rajlater}.

In deep inelastic scattering of an incident lepton off a hadron or nucleus, 
the kinematic invariants are the square of
the momentum carried by the ``space--like'' virtual photon $q^2 = -Q^2 <0$, 
(note: we use the ` -2 ' convention here) and
$x_{Bj} = {Q^2\over 2P\cdot q}$, where $P^\mu$ is the four--momentum of the
target. The cross--section expressed in terms of these invariants is a
product of the point particle Rutherford cross section times a form factor, 
the electromagnetic form factor of the hadron $F_2$. In general, $F_2\equiv
F_2(x_{Bj},Q^2)$, but in QCD, as $Q^2\rightarrow \infty$, $F_2(x_{Bj},Q^2)
\rightarrow F_2(x_{Bj})$. The scaling of the structure function as a 
function of $x_{Bj}$ is what is known as Bjorken scaling. In this lecture, 
we will derive Bjorken scaling using the free field commutation relations.

The cross section for the inclusive deep inelastic scattering process
 $l(k)+(h,A)(P)\rightarrow l (k^\prime) + X$, where $X$ denotes undetected 
final states, is a tensor product of
the leptonic tensor $l^{\mu\nu}$ and the hadronic tensor $W_{\mu\nu}$. The
hadronic tensor is defined as~\cite{Pokorski}
\be
W_{\mu\nu} (q^2,P\cdot q) &=& \sum_{n} (2\pi)^4 \delta^{(4)}(q+P-p_n) 
<P|J_\mu (0) |n><n|J_\nu (0)|P> \, \nonumber \\
&\rightarrow& \int d^4 x\, e^{iq\cdot x} <P|J_\mu (x) J_\nu (0)|P> \, .
\ee
The sum above is over all hadronic final states with momenta $p_n$.  Since 
$q^0 + P^0$ and $p_n^0$ are +ve, we can write the above as
\be
W_{\mu\nu} = \int d^4 x\, e^{iq\cdot x} <P|[J_\mu (x),J_\nu (0)]|P> \, .
\ee
Since the commutator vanishes outside the forward light cone, we will write
the above as
\be
W_{\mu\nu} = \int dx^- e^{iq^+ x^-} \int dx^+ e^{iq^- x^+} \int_{x_t^2 < 2x^+ 
x^-} d^2 x_t <P|[J_\mu (x),J_\nu (0)]|P> \, .
\label{current}
\ee
Above, $J_{\mu}^a = \psibar \gamma_\mu {\lambda^a} \psi (x)$.

In the high energy limit $q^+\rightarrow \infty$, $q^- =$ fixed, the largest
contribution to $W_{\mu\nu}$ comes from the region of the integral 
with the smallest 
oscillations, or $x^+$ finite, $x^- \rightarrow 0$. Since causality 
demands that $x^2 = 2x^+ x^- -x_t^2 < 2 x^+ x^-$, the largest contribution 
to $W_{\mu\nu}$ is from the region of the light cone $x^2 \rightarrow 0$. 
In other words, the structure function is dominated by the light cone 
singularities of the commutator of currents. The limit 
$q^+\rightarrow \infty$ and $q^- =$ fixed, corresponds to the limit 
$\nu = P\cdot q/M
\rightarrow \infty$, $Q^2\rightarrow \infty$ and 
$x_{Bj} = {Q^2\over {2P\cdot q}}$ fixed.

Let us examine the commutator in Eq.~\ref{current} in the limit 
$x^2\rightarrow 0$. Here, using the ``free field'' current commutation relation
which is reasonable in the weak coupling limit,
\be
\{\psibar (x),\psi(-x)\} = {1\over {8\pi}} \gamma^\mu \partial_\mu 
\epsilon (x^0) \delta(x^2) + O(M^2 x^2) \, ,
\ee
we obtain 
\be
[J_\mu (x), J_\nu (-x)] \approx [\psibar (x)\gamma_\mu \gamma_\alpha 
\gamma_\nu \psi (-x) - \psibar (-x) \gamma_\nu \gamma_\alpha \gamma_\mu
\psi (x) ] {1\over {8\pi}}\partial^\alpha \epsilon (x^0) \delta(x^2)
\, .
\ee
We now use the identity
\be
\gamma_\mu \gamma_\alpha \gamma_\nu = S_{\mu\nu\alpha\beta}\gamma^\beta
+ i\epsilon_{\mu\nu\alpha\beta}\gamma^\beta \gamma^5 \, ,
\ee
where
\be
S_{\mu\nu\alpha\beta} = \left(g_{\mu\alpha} g_{\nu\beta} + 
g_{\nu\alpha} g_{\mu\beta} - g_{\mu\nu} g_{\alpha\beta}\right) \, ,
\label{symmtensor}
\ee
and $\epsilon_{\mu\nu\alpha\beta}$ is the anti--symmetric Levi--Civita
tensor in four dimensions.

Substituting this identity in the current commutator, we obtain
\be
& &[J_\mu (x), J_\nu (-x)] \stackrel{x^2\rightarrow 0}{\longrightarrow} 
\Bigg[ \psibar (x) S_{\mu\nu\alpha\beta}\gamma^\beta\psi (-x) 
+ i\epsilon_{\mu\nu\alpha\beta}\psibar (x) \gamma^\beta \gamma^5 \psi (-x)
\nonumber \\
&-& \psibar (-x) S_{\nu\mu\alpha\beta}\gamma^\beta\psi (x) 
- i\epsilon_{\nu\mu\alpha\beta}\psibar (-x) \gamma^\beta \gamma^5 \psi (x)
\Bigg] {1\over {8\pi}} \partial^\alpha \epsilon(x^0)\delta(x^2) \, .
\ee
We now perform a Taylor expansion on $\psi$ and $\psibar$, 
\be
\psibar (x) \psi(-x) = \sum_n {1\over {n!}}\,x^{\mu_1}\cdots x^{\mu_n}
\psibar (0)\stackrel{\longleftrightarrow}{\partial_{\mu_1}}\cdots
\stackrel{\longleftrightarrow}{\partial_{\mu_n}}\psi (0) \, .
\ee
Putting this back into our expression for the commutator, we obtain
\be
[J_\mu (x), J_\nu (-x)] = \sum_{n=1,3}^\infty {1\over  {n!}} 
x^{\mu_1}\cdots x^{\mu_n} {\bf O}_{\beta \mu_1, \cdots ,\mu_n}^{(n+1)} (0) 
S_{\mu\nu\alpha\beta}{1\over {4\pi}} \partial^\alpha \epsilon(x^0)\delta(x^2)
\, ,
\label{intcomm}
\ee
where 
\be
{\bf O}_{\beta \mu_1,\cdots, \mu_n}^{(n+1)} (0) = 
\psibar (0)\gamma^\beta \stackrel{\longleftrightarrow}{\partial_{\mu_1}}\cdots
\stackrel{\longleftrightarrow}{\partial_{\mu_n}}\psi (0) \, .
\label{operator}
\ee
We may note the following points regarding the above result.
\begin{itemize}

\item
Only the odd terms in the sum survive. The even terms cancel out.

\item
We have expanded the operators in the vicinity of the light cone in a 
series of local operators--each of which multiplies the same singular 
function.

\item
Only a particular combination of Lorentz indices appears. We are interested
only in the parity conserving terms, which is why the terms multiplying
the anti--symmetric tensor $\epsilon_{\mu\nu\alpha\beta}$ do not appear. 
In general however, there will be an additional piece proportional to 
$\epsilon_{\mu\nu\alpha\beta}$ which contributes to $W_{\mu\nu}$. The
corresponding structure function often referred to as $F_3$ is measured by
parity violating currents, as for example is the case in deep inelastic 
neutrino scattering.

\item
{\bf O} is a twist two operator. Twist is a term which refers to the 
`dimension' - `spin' of an operator. Our operator above has dimension 
$= 3/2 \times 2 + n$ and spin $= n+1$.
In general, the expansion of the 
operators on the light cone can be organized into an expansion over 
successively higher twists, called the operator product expansion (often known
by its acronym OPE) the coefficients of higher twist operators 
being suppressed by powers of $x^2$. The dominant operators at short 
distances are those with the smallest twist. There are a finite number of
twist two operators.
\end{itemize}

In general, the naive dimensions of the operators are modified by interactions
and they acquire `anomalous dimensions', which may be determined by a 
renormalization group analysis. We will not discuss the OPE any further, but
refer the reader to some of the textbooks with excellent discussions of the 
topic~\cite{Gross,Pokorski,Sterman}.

We return from this digression to topic of immediate interest: 
the derivation of Bjorken scaling. Recall that we had
\be
W_{\mu\nu} = \int d^4 y \,e^{iq\cdot y} <P|[J_\mu (y), J_\nu (-y)]|P> \, .
\nonumber
\ee
We now substitute Eq.~\ref{intcomm} in the RHS of the above. The matrix 
element of the symmetric, traceless operator ${\bf O}^{(n+1)}$ between the
hadronic states, has the tensorial structure, 
\be
<P|{\bf O}_{\beta\mu_1,\cdots ,\mu_n}^{(n+1)} (0)|P> &=&{\bf A}_{n+1}
\,p_\beta 
p_{\mu_1}\cdots p_{\mu_n} + {\bf B}_{n+1}\,
\delta_{\mu_1\mu_2}p_\beta p_{\mu_3}\cdots p_{\mu_n} \nonumber \\
&+& \mbox{less singular terms} \, .
\label{traceless}
\ee
The second term above gives an additional power of $x^2$ when contracted with 
the coefficients and is therefore suppressed. The leading contribution then
is 
\be
W_{\mu\nu} = \int d^4 y\, e^{iq\cdot y} \sum_{n=1,3}^\infty {(p\cdot y)^n 
\over {n!}} {\bf A}_{n+1} S_{\mu\nu\alpha\beta}{1\over {4\pi}} p^\beta 
\partial^\alpha \epsilon(x^0) \delta(x^2) \, .
\ee
Define a function and its Fourier transform
\be
{\tilde f}(p\cdot y) = \sum_{n=1,3}^\infty {(p\cdot y)^n \over {n!}} 
{\bf A}_{n+1}
=\int {dx\over {2\pi}}\, e^{ix y\cdot p}\, {f(x)\over x} \, .
\ee
Substituting the above into $W_{\mu\nu}$ and using the identity
\be
\int d^4 y\, e^{iky} \delta(y^2)\epsilon(y^0) = (2\pi)^2 \epsilon(k^0)\delta(k^2)
\, ,
\ee
we obtain
\be
W_{\mu\nu} = \int {dx\over {2\pi}} {f(x)\over x} p^\beta (q+xP)^\alpha 
S_{\mu\nu\alpha\beta} (2\pi)^2 \epsilon(xP^0 + q^0) {1\over 4\pi} 
\delta( (xP+q)^2) \, .
\ee
Using the definition of $S_{\mu\nu\alpha\beta}$ in Eq.~\ref{symmtensor} and
performing the delta function integration which sets $x\equiv x_{Bj} =
-q^2/2P\cdot q$, we can write the above finally as
\be
W_{\mu\nu} = {f(x)\over {(P\cdot q)}} \left(P_\mu - {(P\cdot q) q_\mu 
\over { q^2}}\right)\left(P_\nu - {(P\cdot q) q_\nu\over {q^2}}\right)
-{f(x)\over {2x}} \left(g_{\mu\nu}- {q_\mu q_\nu \over {q^2}}\right) \, .
\label{Bj}
\ee

The electromagnetic tensor $W_{\mu\nu}$ has the most general
tensorial decomposition,
\be
W_{\mu\nu} = a_1 P_\mu P_\nu + a_2 P_\mu q_\nu + a_3 P_\nu q_\mu +
a_4 q_\mu q_\nu + a_5 g_{\mu\nu} \, . \nonumber
\ee
The symmetry properties require however that $a_2 = a_3$ and from current
conservation $q^\mu W_{\mu\nu}=0$, and similarly for $q^\nu W_{\mu\nu}=0$, 
we obtain,
\be
W_{\mu\nu} = 
{F_2\over {(p\cdot q)}} \left(p_\mu - {(p\cdot q) q_\mu 
\over {q^2}}\right)\left(p_\nu - {(p\cdot q) q_\nu\over {q^2}}\right)
-F_1 \left(g_{\mu\nu}- {q_\mu q_\nu \over {q^2}}\right) \, ,
\ee
where $F_1$ and $F_2$ are the structure functions. Comparing the above 
to our result Eq.~\ref{Bj}, we observe that $F_2 = f(x_{Bj})$, which is the
famous scaling phenomenon known as Bjorken scaling. Further, in this limit
$F_1 = F_2/2x$--this result is known as the Callan--Gross relation.

We shall now show that the structure functions derived above can be
simply related, in leading twist, to the light cone parton distributions and 
further show that $F_2$ thereby 
has the intepretation of being the probability that a quark has a fraction 
x of the total hadron momentum $p^+$ on the light front.

Consider the forward Compton scattering amplitude for the virtual photon
scattering of the hadron in deep inelastic scattering,
\be
T_{\mu\nu} (q^2,p\cdot q) = i\int d^4 z e^{iq\cdot z} <P\mid T(J_\mu (z) 
J_\nu (0))\mid P> \equiv 2 {\rm Im} W_{\mu\nu}\, .
\ee
This can be decomposed into longitudinal and transverse pieces
\be
T_{\mu\nu} = {p_\mu p_\nu \over {M^2}}\,t_{\perp} (x,q^2) -
g_{\mu\nu}\, t_L(x,q^2) \,,
\ee
just as for $W_{\mu\nu}$. Now, in the Bjorken limit, the Callan--Gross 
relation implies that the longitudinal piece above vanishes. To leading 
twist then, just as for the hadronic tensor, we can decompose the 
transverse component of the Compton amplitude as 
\be
t_{\perp}^{T=2} = \sum_{n=1}^{\infty} \int d^4 z \,e^{iq\cdot z} 
C_n^\beta (z^2)\, z^{\mu_1,\cdots ,\mu_n}
<P\mid {\bf O}_{\beta\mu_1,\cdots, \mu_n}\mid P> \, ,
\ee  
where, making the analogy to Eq.~\ref{intcomm}, the coefficient functions 
$C_n^\beta (z^2)$ are the same for all odd values of $n$ and 
zero otherwise. Also, ${\bf O}$ is the operator defined in
Eq.~\ref{operator}~\footnote{In general, the partial derivatives in Eq.~\ref
{operator} should be replaced by covariant derivatives.}. One can define
\be
{2^{n} q^{\mu_1}\cdots q^{\mu_n} \over {(-q^2)^{n+1}}} {\tilde C}_n^\beta (q^2)
= i\int d^4 z \exp(iq\cdot z) z^{\mu_1}\cdots z^{\mu_n} C_n^\beta (z^2) 
\, .
\ee
Typically, the functions ${\tilde C}_n^\beta (q^2)$ are different and are
the coefficient functions in the operator product expansion. However, in the 
scaling limit, they are constants. Substituting the above
identity into our expression for $t_{\perp}^{T=2}$, we obtain
\be
{p\cdot q \over M^2}\,t_{\perp}^{T=2} (x,q^2) = {-2q^2\over {p\cdot q}} 
\sum_{n=1,3}^\infty \left(2q_{\beta}\over q^2 \right)\left(2q_{\mu_1}\over 
q^2\right)\cdots 
\left(2q_{\mu_{n}} \over q^2\right) <P\mid {\bf O}^{\beta \mu_1\cdots \mu_n}
\mid P> \, .
\ee

Since ${\bf O}$ is traceless and symmetric, we can again use the tensorial
decomposition in Eq.~\ref{traceless}. Then, since $x= -q^2/2p\cdot q$, 
we obtain
\be
{p\cdot q \over M^2} t_\perp^{T=2} = 4 x \sum_{n=1,3}^\infty \left(
{-1\over x}\right)^
{n+1} {\bf A}_{n+1} \, .
\ee
We can determine ${\bf A}_{n+1}$ by setting all the Lorentz indices in
Eq.~\ref{traceless} to $+$. Then,
\be
{\bf A}_{n+1} = \left({1\over {p^+}}\right)^{n+1} <P\mid {\bf O}^{++\cdots +}
\mid P>_C \, .
\label{operA}
\ee
From the definition of the operator ${\bf O}$ in Eq.~\ref{operator}, the 
matrix element above is given, in light cone gauge $A^+ =0$, 
by all two particle irreducible insertions of
the vertex $\psibar \gamma^+ (k^+)^n \psi$~(see Ref.~\cite{Jaffe} and
references therein).

Let us now digress a little to discuss the light cone Fock space
distribution. We will relate it subsequently to the structure functions
above. Recall the decomposition we had in Eq.~\ref{normalize} of 
lecture 1 of the dynamical 2--spinor $\psi_+$. We can then define the
light cone parton distribution function as
\be
{dN\over d^3 k} = {1\over (2\pi)^3}  \sum_\lambda \left[b_\lambda^\dagger
b_\lambda + d_\lambda^\dagger d_\lambda \right] \, .
\ee
Writing this in terms of $\psi_+$ and using the light cone identity
\be
{\rm Tr}\left[ \sum_\lambda \gamma^+ \psi_\lambda (x)\psibar (y)\right]
= \sqrt{2}
{\rm Tr} \left[\sum_\lambda \psi_{+,\lambda}(x)\psi_{+,\lambda}^\dagger (y)
\right] \, ,
\ee
we obtain
\be
{dN\over d^3 k} = {2\over {(2\pi)^3}}\int d^3 x d^3 y e^{-ik\cdot (x-y)}
{\rm Tr} \left[\gamma^+ S(x,y)\right] \, ,
\ee
where $S(x,y)=-i<T(\psi(x)\psibar(y))>$. The light cone distribution function
integrated over all momenta is the function
\be
H(\alpha) = \int{d^4 k \over {(2\pi)^4}}\, \delta(\alpha - {k^+\over p^+}) 
{1\over p^+} {\rm Tr} \left[\gamma^+ {\tilde S} (p,k)\right] \, ,
\label{funcH}
\ee
where ${\tilde S}(p,k)$ is the fermion Green's function in momentum space.
We will now show that the function $H(\alpha)$ is, in leading twist, the
structure function $F_2$.

Returning now to Eq.~\ref{operA}, we find
\be
{\bf A}_{n+1} = {1\over {(p^+)^{n+1}}} \int {d^4 k \over {(2\pi)^4}} 
(k^+)^n {\rm Tr} \left[\gamma^+ {\tilde S}(p,k)\right] \, .
\ee
In terms of $H(\alpha)$ then, 
\be
{\bf A}_{n+1} = \int_{-\infty}^\infty d\alpha \,\alpha^n H(\alpha) \, .
\ee
From the analytic properties of the function $H(\alpha)$, specifically the
anti--commutation properties of the operators $\psibar \gamma^+$
and $\psi$ on the light cone~\cite{Jaffe}, 
one may conclude that $H(\alpha)= 0$ for
$|\alpha|>1$. Substituting the expression for ${\bf A}_{n+1}$ in the 
transverse Compton amplitude, we obtain,
\be
{p\cdot q\over M^2}\, t_\perp^{T=2} = 4\int_{-1}^1 d\alpha \sum_{n=1,3}^\infty
\left({\alpha\over x}\right)^n H(\alpha) \, .
\ee
Peforming the sum over $n$ and analytically continuing $t_\perp$ to the
physical region $x\rightarrow x-i\epsilon$, with x real and $0<x\leq 1$, 
\be
{p\cdot q\over M^2}\,t_\perp^{T=2} = 2x\int_{-1}^{1} d\alpha \,H(\alpha) 
\left\{ {1\over {x-\alpha -i\epsilon}}-{1\over {x+\alpha-i\epsilon}}\right\}
\, .
\ee
Taking the imaginary part of the amplitude to obtain the structure 
functions, we get
\be
F_2(x) = x(H(x)-H(-x)) \, .
\label{qpmodel}
\ee

From the definition of $H(x)$ in Eq.~\ref{funcH}, it is the probability to
find a quark with momentum $k^+ = x p^+$ in the target. The function 
$-H(-x)$ has the interpretation of finding an anti--quark with momentum 
$k^+ = x p^+$ in the target. We have therefore, with Eq.~\ref{qpmodel},  
obtained the usual parton model interpretation of structure functions.  In 
general, for a large but finite $Q^2$, the above result can be slightly 
modified to read
\be
F_2(x,Q^2) = \int_0^{Q^2} d^2 k_t {dN\over {d^2 k_t dx}} \, .
\ee
This follows simply from putting an upper cut--off $Q^2$ on the $k_t$ 
integration in Eq.~\ref{funcH}. Finally, we should mention that 
the multi--parton Fock distributions discussed in lecture 3 can be
related by a similar analysis to the higher twist  contributions to the 
forward Compton scattering amplitude~\cite{Jaffe, EllisFurmPet}.

\section*{Acknowledgements}
I would like to thank the organizers of the Eleventh Chris Engelbrecht Summer 
School in Theoretical Physics for inviting me deliver these lectures in 
Cape Town, SA. In particular, I would like to thank Prof. Jean Cleymans for
his gracious hospitality. I would also like to thank the students and other
participants at the school who helped create a stimulating environment. This 
work was supported by the Danish Research Council and the Niels Bohr Institute.

\end{document}